# Optimal Routes to Ultrafast Polarization Reversal in Ferroelectric LiNbO$_3$

R. Tanner Hardy[1], Conrad Rosenbrock[2], Gus L. W. Hart[2], and Jeremy A. Johnson[1]

[1] Department of Chemistry and Biochemistry, Brigham Young University, Provo, Utah, USA 84602
[2] Department of Physics and Astronomy, Brigham Young University, Provo, Utah, USA 84602

Electronic Mail: jjohnson@chem.byu.edu, gus_hart@byu.edu

**Abstract:**
   We use the frozen phonon method to calculate the anharmonic potential energy surface and to model the ultrafast ferroelectric polarization reversal in LiNbO$_3$ driven by intense pulses of THz light. Before stable switching of the polarization occurs, there exists a region of excitation field-strengths where transient switching can occur, as observed experimentally [Physical Review Letters 118, 197601 (2017)]. By varying the excitation frequency from 4 to 20 THz, our modeling suggests that more efficient, permanent polarization switching can occur by directly exciting the soft mode at 7 THz, compared to nonlinear phononic-induced switching driven by exciting a high frequency mode at 18 THz. We also show that neglecting anharmonic coupling pathways in the modeled experiment can lead to significant differences in the modeled switching field strengths.

   A crucial capability to develop high-speed electronic devices is the ability to control atomic structure on picosecond timescales with pulses of light [1,2]. For example, atomic-structure based switches could be developed using ferroelectric materials where structural changes lead to states that are equivalent in energy, but manifest macroscopic polarization that points in opposite directions [3]. These different polarization states can have large enough potential energy barriers as to prevent thermal hopping over the barrier, providing a long-term stable "bit". Intensive research has been focused on ultrafast light-induced switching of the polarization because theoretical studies have shown picosecond switching times that would enable disruptive advances in the operation speed of non-volatile memory [3-6].
   Different experimental routes have been suggested and tested to switch the macroscopic polarization of ferroelectric materials by directly or indirectly driving the zone-center soft mode (the phonon mode with the strongest ties to the polarization). Coherent control methods, also recently termed *nonlinear phononics* [3,4,7-12], involve the simultaneous strong excitation of different atomic motions in the material. In theory, these approaches can lower the potential energy barrier and ease the switching of ferroelectric polarization.
   Recently, Mankowsky and co-workers showed, in a somewhat surprising manner, they could *transiently* switch the ferroelectric polarization in LiNbO$_3$ [10]. With intense pulses of 19 THz light, they resonantly excited the highest frequency A$_1$ symmetry mode and used a second-harmonic generation (SHG) probe to monitor the sample (see Figs. 1 and 2 in Ref. [10]). The SHG intensity is proportional to the ferroelectric polarization and therefore sensitive to the distance of the Li atoms from the midpoint between two Nb atoms. When LiNbO$_3$ was excited with high enough fluence, the second harmonic probe intensity was reduced to zero, recovered briefly, then returned to zero intensity, and recovered to the initial intensity in ~1 ps. They interpreted the signal as resulting from the low frequency (soft) A$_1$ mode being anharmonically driven to such large amplitudes that the Li atoms cross the midpoint between Nb atoms, making it transiently centrosymmetric and stopping SHG. The Li atoms travel some distance on the other side, slightly recovering the SHG with the ferroelectric polarization (and phase of the SHG light) switched.

But surprisingly, the atoms then rebound (crossing the midpoint again) and return to their initial equilibrium positions as the motion is damped on the 1 picosecond timescale. This transient polarization switching was later observed again with an alternative femtosecond stimulated Raman scattering probing scheme [13].

Mankowsky's experiment attempted to verify a promising computational prediction by Subedi [6], who suggested that the ultrafast reversal of ferroelectric polarization could be accomplished by exciting a high frequency mode that couples anharmonically to the soft $A_1$ mode. Surprisingly, the experiment showed only temporary polarization reversal, suggesting that the atoms switched and then returned to their original positions.

In this letter, we present first-principles calculations that confirm the possibility of a transient switching of the ferroelectric polarization upon intense THz excitation. We calculated the interatomic potential energy surface (PES), including all of the symmetry-allowed $A_1$ modes for $LiNbO_3$, and then we ran a virtual experiment parallel to the experiments performed in Ref. [10].

To perform these virtual experiments, we fully relaxed the $LiNbO_3$ structure, we calculated frequencies and eigenvectors of zone-center phonon modes, then we displaced along pairs of phonon modes to reconstruct the anharmonic potential energy surface and extract coupling constants. The structure was relaxed until the force on each atom was less than $4 \times 10^{-4}\ eV/Å$ ($10^{-5}\ Hartree/Bohr$) and stress on the unit cell was less than 1 kb ($3 \times 10^{-6}\ Hartree/Bohr^3$). Phonon eigenvectors and frequencies were calculated using the DFT-GGA functional as used by Perdew-Burke-Ernzerhof [14], using a plane-wave basis set and Projector Augmented Wave Pseudopotentials (POT-PAW), as implemented in the VASP code [15-19]. By displacing each atom of $LiNbO_3$, we calculated the forces on the rest of the atoms to determine the dynamical force matrix. We then used the PHONOPY code [20] to calculate the frequencies and eigenvectors for all the modes (see Fig. 1(a)). We used a $6 \times 6 \times 12$ k-point grid for Brillouin zone integration and a 700 eV cutoff energy for the plane-wave basis expansion.

$LiNbO_3$ has four phonon modes with $A_1$ irreducible representations and zone-center frequencies of 7 THz (the "soft" mode), 7.4 THz, 9.0 THz, and 18.3 THz (the 18.3 THz mode was excited in the Refs. [10,13] experiments). Eigenvectors for these modes are shown in Fig. 1(a). We used the frozen phonon method [6,8] to calculate the potential energy surfaces for each pair of the four $A_1$ zone-center phonon coordinates (six pairwise potential energy surfaces in total). We then fit the calculated potentials to the two-mode potential energy equation (Eq. 1), to obtain the nonlinear coupling coefficients up to 4$^{th}$ order.

$$V(Q_i, Q_j) = \frac{1}{2}\omega_i^2 Q_i^2 + \frac{1}{2}\omega_j^2 Q_j^2 + \sum_{ij} c_{ijk} Q_i Q_j Q_k + \sum_{ij} d_{ijkl} Q_i Q_j Q_k Q_l \qquad (Eq.\ 1)$$

$\omega_i$ is the harmonic angular frequency of each vibrational mode, $c_{ijk}$ are the third-order coupling constants, and $d_{ijkl}$ are the fourth order coupling constants. Note that *k* and *l* in the sums will only take on values of *i* and *j*, because only a pair of modes were used to calculate each surface.

To model the excitation and the resulting polarization dynamics of this system, we treated all four modes as coupled classical underdamped oscillators, experiencing the external driving force due to the electric field of an ultrafast THz pulse. The coupling constants were determined directly from the fits of our six pair-wise potential energy surfaces, and we use experimentally determined linear damping rates for each phonon mode.

We ran our virtual experiment by exciting the system with a ~100 fs THz pulse with a central frequency that we vary from ~4 to 20 THz. For each central frequency, we then increase the strength of the THz driving field until switching the polarization occurs. Fig. 1(b) shows the modeled time-dependent ferroelectric polarization for two driving field magnitudes, one below the switching threshold, and one above (showing that the polarization switches direction and stays switched). Figure 1(c) is helpful to consider in understanding the ultrafast atomic motion and polarization switching dynamics. The contour lines in Fig. 1(c) show the potential energy surface determined from displacing the soft $A_1$ mode at 7 THz and the high frequency $A_1$ mode at 18 THz. Two local minima are apparent in the PES. The system originates from the minimum on the right, and if the driving field is strong enough to overcome the potential barrier with the correct trajectory, the system will relax to the minimum on the left (corresponding to a reversal of the ferroelectric polarization). The false color map indicates the magnitude of the ferroelectric polarization, with negative (blue) values on the right, and positive (red) values on the left.

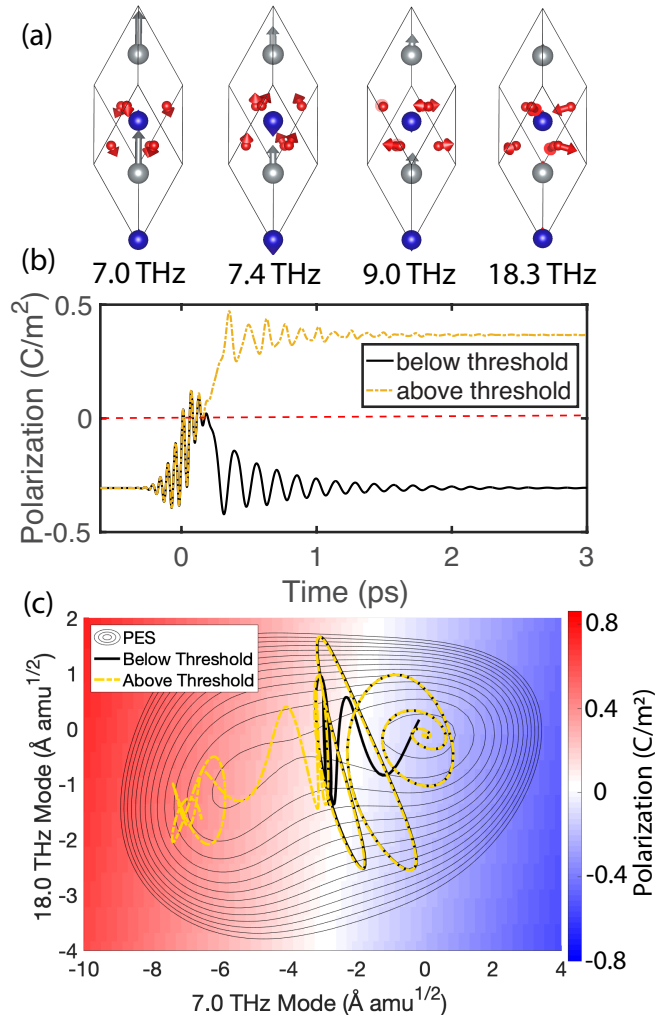

Figure 1. (a) Frequencies and eigenvectors of four $A_1$ phonon modes in $LiNbO_3$. (b) Modeled time-dependent polarization after multi-THz excitation. Black line at lower pump powers below the switching threshold and yellow-orange dash-dotted line above the switching threshold. (c) Contour plot of PES when displacing along 7 and 18 THz $A_1$ mode coordinates. The black and yellow-orange lines are trajectories for polarization plotted in (b). The false-color scale indicates the polarization for atomic coordinates on the surface.

The orange and black lines show the trajectories for a given THz driving field for the first 0.7 ps, in this case with driving frequency centered at 18 THz. Immediately, coupling between modes during the 100 fs driving pulse results in motion along both phonon coordinates. If the THz pulse is strong enough, the system crosses the potential barrier, permanently relaxing to the minimum with opposite polarization (orange dash-dotted line). If the electric field is not strong enough, the system will always relax back to the starting position (black solid line). Note that even though we plot trajectories only along these two phonon coordinates, the anharmonic coupling calculated from other pairwise potentials between all $A_1$ phonon coordinates are included in the modeling.

The time-dependent polarization shown in Fig. 1(b) was calculated using the trajectories shown in Fig. 1(c), confirming that a transient switching of ferroelectric polarization is possible; we see that the polarization indicated by the black line crosses the zero point temporarily into the red-shaded region, but then returns to the original static polarization value. A careful inspection of Fig. 1(c) reveals that the point where the magnitude of the ferroelectric polarization goes to zero does not line up exactly with the potential energy barrier, making a transient crossing of the zero-polarization point possible without a permanent switching.

Ref. [10] discussed the possibility that other modes or neighboring ferroelectric domains with opposite polarization could have played a role in the observed transient switching dynamics, which are certainly possibilities. But even in our model that neglects those additional affects, we observe that transient switching of the polarization is possible, and it occurs on identical time scales to what was seen in experiments.

With good agreement to experimental data to validate our approach, we can also use this computational approach to explore other potential methods to accomplish permanent ultrafast polarization switching. For example, we can sweep the central frequency of the pump pulse and determine a relative field strength required to switch the ferroelectric polarization for each pump frequency. Figure 2 shows the relative switching field as a function of central pump frequency. The different blue dashed lines correspond to unique carrier-envelope phases of the pump pulse, and the average switching field is the solid black line.

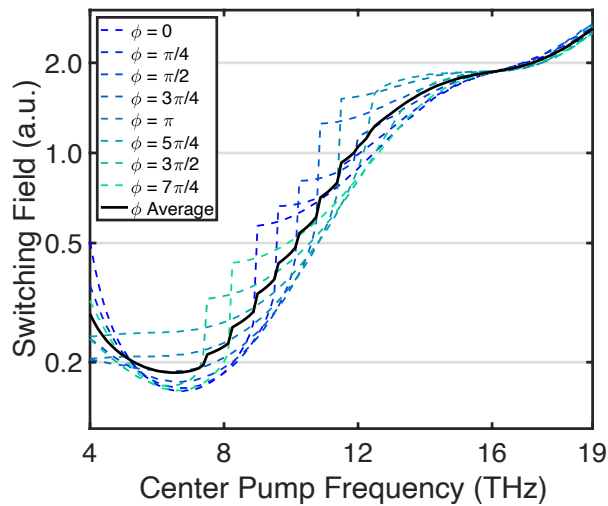

Figure 2. Relative peak electric field required to switch the ferroelectric polarization for specific pump central frequencies. The blue dashed lines indicate specific carrier-envelope phases of the pump pulse, and the solid black line is the average.

We see that the lowest switching field is not close in frequency to the 18 THz mode, but resonant with the lowest frequency 7 THz mode. This mode is the soft mode tied to the ferroelectric polarization, and we see that directly driving the soft mode is predicted to be the most efficient route to polarization switching, with a ~10× lower switching field than required when utilizing anharmonic coupling of the high frequency mode. This suggests that directly pumping the soft mode should be the most efficient route to ferroelectric polarization reversal in $LiNbO_3$.

We can also compare our four-mode model with a simpler model similar to previous work in Ref. [6] that only included two modes in the calculations: the high-frequency 18 THz mode and the 7 THz soft mode. We compute the average switching field as a function of frequency for both models. Figure 3 shows the percent difference in switching field comparing the two models for $LiNbO_3$. A value larger than zero indicates that the two-mode model predicts a higher switching field than the four-mode model, and a value less than zero indicates the four-mode model predicts a lower switching field. We see that at high frequencies, the two-mode model underestimates the switching field. This difference arises from the fact that the other two $A_1$ phonon modes at 7.4 THz and 9 THz also couple to the high frequency phonon, and some energy deposited in the high frequency mode transfers to these two modes instead of the soft mode motion that corresponds to polarization switching.

On the other hand, at lower frequencies from ~5 to 10 THz, the four-mode model predicts a lower switching field than the two-mode model. This lower predicted field arises due to the moderate bandwidth of the pump pulse. With the frequency centered below 8 THz, the THz pulse resonantly drives not only soft mode, but also the other two low-frequency modes. We directly put energy into the soft mode with the driving field, but then additional energy can flow to the soft mode via coupling with the other two resonantly excited modes. Both of these cases (excitation at higher and lower frequencies) indicate that the anharmonic interactions to other modes with symmetry-allowed couplings should not be neglected when trying to predict ultrafast dynamics.

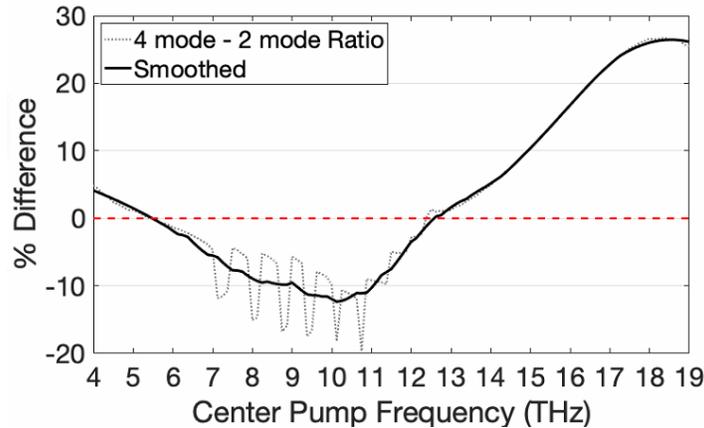

Figure 3. Ratio of peak switching field of the four-mode model compared to the two-mode model, showing the two-mode model does not accurately capture the switching dynamics at low and high frequencies.

Finally, with our more complete four-mode model that shows good agreement to experimental data, we can also investigate methods to tailor the excitation pulses to reverse the ferroelectric polarization more efficiently. For example, we hypothesize that a linearly chirped pump pulse with lower frequencies within the pump bandwidth arriving later could more efficiently drive the motion as the PES softens going across the potential barrier between polarization states.

In conclusion, we have used the frozen phonon method to model ferroelectric polarization switching in LiNbO$_3$. We find that under certain conditions, we can reproduce with nearly identical temporal dynamics the transient switching of polarization observed by Mankowsky et al. [10]. We show that neglecting symmetry allowed anharmonic couplings to other modes in the modeling can lead to significant differences in predicted switching field strengths. Furthermore, we vary the excitation frequency to show that significantly lower switching fields are predicted to occur when resonantly driving the soft mode, rather than driving a higher frequency mode that couples to the soft mode. The idea that directly exciting the soft mode to achieve polarization switching should warrant more investigation. Particularly due to reports of using meta-material layers or antennas to locally enhance THz electric fields [21-28]; these local enhancement methods work better with lower frequencies due to larger feature sizes and easier fabrication.

enhancements. Opt. Express **20**, 8551 (2012). http://www.opticsexpress.org/abstract.cfm?URI=oe-20-8-8551